\documentclass[preprint,showpacs,aps]{revtex4}

\usepackage[dvips]{graphicx}
\usepackage{color}
\usepackage{amssymb}
\usepackage{amsmath}
\usepackage{units}
\usepackage{rotating}
\usepackage[latin1]{inputenc} 

\newcommand{\FeCo}{Fe$_x$Co$_{100-x}$}
\newcommand{\CFS}{Co$_2$FeSi}
\newcommand{\CMS}{Co$_2$MnSi}
\newcommand{\CFA}{Co$_2$FeAl}
\newcommand{\CMA}{Co$_2$MnAl}
\newcommand{\CMG}{Co$_2$MnGe}
\newcommand{\CYZ}{Co$_2$\textsl{XZ}}
\newcommand{\Nv}{N_\mathrm{v}}

\begin{document}

\title{Exchange stiffness in Co$_{2}$-based Heusler compounds}

\author{Jaroslav Hamrle,$^{*}$ Oksana Gaier, Simon Trudel, Burkard Hillebrands}

\affiliation{Fachbereich Physik and Forschungszentrum OPTIMAS,
Technische Universit\"at Kaiserslautern,
Erwin-Schr\"odinger-Stra\ss e 56, D-67663 Kaiserslautern, Germany}

\author{Horst Schneider, Gerhard Jakob}
\affiliation{Institut f\"ur Physik,
Johannes-Gutenberg-Universit\"at Mainz, Staudinger Weg 7, D-55128
Mainz, Germany}

\begin{abstract}
We determine the spin wave exchange stiffness $D$ and the exchange constant $A$ for the full Heusler compound \CFS\ using Brillouin light scattering spectroscopy. We find an extraordinarily large value of $D=715\pm20$\,meV\,\AA$^2$ ($A=31.5\pm1.0$\,pJ/m) which is, to the best of our knowledge, only surpassed by the intermetallic compound Fe$_{53}$Co$_{47}$ (J. Appl. Phys. \textbf{75}, 7021 (1994)). 
Furthermore, we provide a systematization of the exchange stiffnesses determined for a variety of Co$_2$-based Heusler compounds. We find that for the  investigated compounds, the exchange stiffness is a function of the valence electron concentration and the crystallographic ordering. The exchange stiffness increases when the valence electron concentration and/or the amount of the L2$_1$ ordering  increase. A qualitative explanation for the dependence on the valence electron concentration is provided.

$^{*}$Corresponding author: J. Hamrle (hamrle@physik.uni-kl.de)
\end{abstract}

\pacs{75.30.Et, 75.30.Ds, 75.50.Cc, 78.35+c}

\maketitle

Heusler compounds, which are predicted to be half-metallic materials with a 100\% spin polarization at the Fermi level, are attracting considerable attention for their use in spintronic devices \cite{sak06,review:Inomata}. 
However, inelastic electron-magnon interactions can create states near the Fermi level in the gap of the minority spin channel, which reduce the idealized 100\% spin polarization \cite{Skomski,rajanikanth:063916}.
As such, the investigation of these phenomena in Heusler compounds is a pressing
issue in order to understand the strong temperature dependence of
the spin polarization of these materials \cite{sak06,rajanikanth:063916}.
In addition to spintronic devices, several proposed applications for which Heusler compounds are promising contenders include 
spin-calorimetric, 
magneto-optical,  
and devices based on ferromagnetic shape memory alloys.
In all of these applications, a key parameter is the exchange stiffness $D$ which describes the energy of a magnon, and is related to the exchange constant $A$ which expresses the energy of aligned spins in a magnetic material. 
The knowledge of these material parameters is also crucial for micromagnetic simulations and the study of dynamic phenomena. While there have been several theoretical investigations of exchange in Heusler compounds \cite{Thoene}, a systematic experimental investigation and comparison of the exchange stiffness in Heusler compounds is still lacking.

Our recent efforts towards this goal have concentrated on the use of Brillouin light scattering (BLS) spectroscopy to determine the exchange stiffnesses of a variety of Co$_2$-based Heusler compounds with composition \CYZ\ \cite{gai08,ham09cms,gai09ccfa}, where \textsl{X} and \textsl{Z} are a transition metal and a main group element, respectively.
In the following, we first present a BLS study of \CFS\ thin films. We emphasize how the value of the exchange stiffness is determined from the measured BLS spectra. In particular, we will show that \CFS\ provides a very high exchange stiffness $D$. To the best of our knowledge, only \FeCo\ alloys where $x\sim$50 are reported to provide higher values \cite{liu94}. 
Thereafter, a systematization of experimentally determined exchange stiffness for \CYZ\ Heusler compounds is provided. The observed trend is parameterized by the number of valence electrons $\Nv$, and a tentative qualitative explanation for this trend is provided.

The \CFS\ samples investigated here consist of Al(\unit[4]{nm})/\CFS($d$)/Cr(\unit[30]{nm})/MgO(001) epitaxial structures ($d$=20 and \unit[60]{nm}),
in which all layers were deposited by pulsed laser deposition using a KrF ($\lambda_{\mathrm{laser}}$ = \unit[248]{nm}, pulse energy \unit[600]{mJ}). 
After deposition the films were annealed at \unit[450]{$^{\circ}$C}. 
This results in L2$_1$ ordered \CFS\ samples,
as is confirmed by the presence of (111)
Bragg reflections in the x-ray diffractograms, which are forbidden for structures with lower atomic ordering, such as the B2 structure. Details on the sample preparation are provided in Ref.~\cite{schneider2009}.

The BLS measurements were performed in the magneto-static surface
mode geometry, \emph{i.e}.\ the magnetic field $H$ was applied parallel
to the film surface and perpendicular to the plane of light
incidence. A diode pumped, frequency doubled Nd:YVO$_4$ laser ($\lambda=532$\,nm) was used as a light source. A description of this BLS setup can be found in Ref.~\cite{hil99}. Unless otherwise indicated, BLS spectra were recorded at an angle of incidence $\varphi=$\unit[45]{$^\circ$}, defined as the angle between the direction of the probing light beam and the film plane's normal. The corresponding in-plane wave vector of the detected magnons is given as 
$q_\|=(4\pi/\lambda)\cdot \sin\varphi$, being $1.67\cdot10^7$\,m$^{-1}$ for $\varphi=$\unit[45]{$^\circ$}. 

\begin{figure}[tbp]
\includegraphics[width=0.43\textwidth]{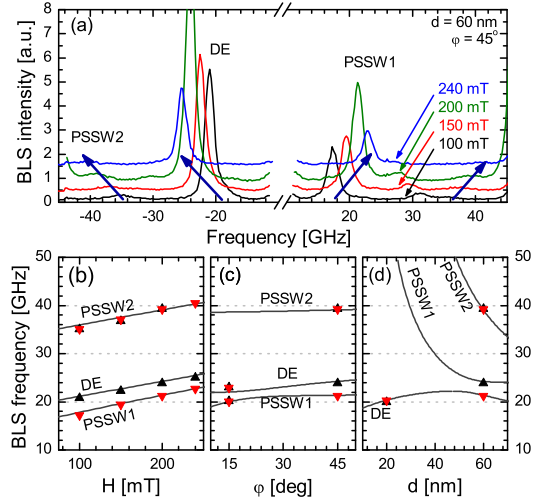} 
\caption{%
\label{fig:BLS}%
(color) (a) BLS spectra for 60~nm thick \CFS\ film in different magnetic fields.
Peaks are assigned as Damon-Eshbach (DE) or perpendicular standing spin wave (PSSW) modes. (b) field dependence of the BLS frequencies ($\blacktriangle$ Stokes lines, \textcolor{red}{$\blacktriangledown$} anti-Stokes). (c) and (d) dependence of the BLS frequencies on the angle of incidence $\varphi$ ($H$ = \unit[200]{mT} and $d$ = \unit[60]{nm}), and the film thickness $d$ ($H$ = \unit[200]{mT}), respectively. (b--d) also show the results of numerical simulations of the BLS frequencies (solid lines). See text for details. 
}
\end{figure}

BLS spectra measured for the 60 nm thick \CFS\ sample in various external magnetic fields are presented in Fig.~\ref{fig:BLS}(a).
The field dependence of the spin wave frequencies $\omega$ is shown in Fig.~\ref{fig:BLS}(b). For both the Stokes (negative frequency) and anti-Stokes (positive frequency) parts of the spectrum, $\omega$ shifts to higher values with increasing magnetic field. 
This is evidence of the magnonic origin of the observed peaks 
\cite{Hilleb-LSiS}
\footnote{%
The peak \emph{ca.} 30\,GHz in Fig.~\ref{fig:BLS}(a) moves to lower $\omega$ with increasing $H$. This is a 2$^{\mathrm{nd}}$ order transmission peak
of the Damon-Eshbach (DE) mode resulting from the finite finesse of our Fabry-P\'{e}rot interferometer. It does not correspond to a real spin-wave mode at a given $\omega$, and is not considered later on. 
}.
The dependence of spin-wave frequencies on $\varphi$ (\emph{i.e.}\ on $q_\|$) and on \CFS\ thickness is shown in Fig.~\ref{fig:BLS}(c) and (d) respectively.

The value of the exchange stiffness $D$ is determined by fitting the experimental spin wave frequencies using a phenomenological model~\cite{hil90}. The exchange constant $A$ is determined from $A=D M_{\mathrm{S}} / 2g \mu_\mathrm{B}$
where $M_\mathrm{S}$ is the saturation magnetization, $g$ is the Land\'{e} $g$-factor and $\mu_\mathrm{B}$ is the Bohr magneton \cite{ham09cms}. The spin wave frequencies were calculated as a function of (a) the external magnetic field, (b) the angle of incidence  and (c) the film thickness (Fig.~\ref{fig:BLS}(b-d)).
For all three dependencies the best agreement between the experimental data  and the calculations is achieved for $D=715\pm20$\,meV\,\AA$^2$ ($A=31.5\pm1.0$\,pJ/m), 
$\mu_0 M_\mathrm{S}=1.28$\,T 
($\mu=4.93\,\mu_\mathrm{B}/$formula unit),
and $g=2.0$, respectively. 
The magnetic anisotropies were neglected, as we have verified  that they are too small  to change the spin-wave frequencies  observed in the BLS spectra (\emph{i.e.} below $10^3$\,J/m$^3$, data not shown). This is in agreement with negligibly small magnetocrystalline anisotropy we have previously found in L2$_1$ ordered Co$_2$MnSi  \cite{gai08}.

The Heusler compounds are well known to be systematized by the number of valence electrons $\Nv$. Striking examples are the magnetic moment described by the Slater-Pauling rule \cite{SP-Galanakis,fech06,kueb84} or the Curie temperature $T_C$ \cite{kub07}, both linearly dependent on $\Nv$. Here we show that the exchange stiffness $D$ in Heusler compounds scales with $\Nv$ as well.
We have collected $D$ values of various Co$_2$-based Heusler compounds. Most of these were determined by us, namely: Co$_2$FeSi(L2$_1$) (this work), Co$_2$MnSi(L2$_1$) \cite{ham09cms}, Co$_2$FeAl(B2) and Co$_2$Cr$_{0.6}$Fe$_{0.4}$Al(CCFA)(B2) \cite{gai09ccfa}, Co$_2$MnSi$_{1-x}$Al$_x$, including Co$_2$MnAl(B2) \cite{kub09}. Belmeguenai \textit{et al.} recently reported on the exchange stiffness of Co$_2$MnGe \cite{bel09}. However, the crystallographic order of the investigated \CMG\ films was not specified. All values of $D$ were determined using BLS spectroscopy measurements carried out at room temperature (RT).

\begin{figure}[tbp]
\includegraphics[width=0.4\textwidth]{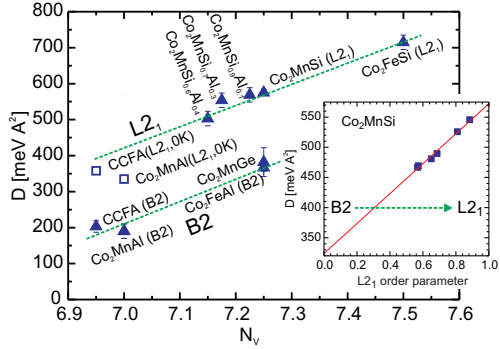}
\caption{%
\label{f:D-Heuslers}
(color)(full triangles) Experimental exchange stiffness $D$ of various \CYZ\ compounds. The straight dashed lines are guide for eye for L2$_1$ and B2 ordered compounds. (empty squares) Expected $D$ values for CCFA and Co$_2$MnAl when corrections for L2$_1$ order and 0\,K are taken into account (see text). Inset: Dependence of $D$ on L2$_1$ order for Co$_2$MnSi. The straight line shows a linear fit. Data determined from our previous work \protect\cite{gai08,ham09cms}. 
}
\end{figure}

The dependence of $D$ on $\Nv$ is presented in Fig.~\ref{f:D-Heuslers}. The salient features are:
(i) A very large change in $D$ is observed between \CFS\ ($D$=715\,meV\,\AA$^2$) and CCFA or \CMA\ (both $\sim$ 200\,meV\,\AA$^2$). Such a large change is associated with the introduction of only 0.5~valence electron per atom.
(ii) The experimental points are segregated into two branches related to B2- and L2$_1$-ordered compounds. In both branches, the $D$ is monotonously increasing with increasing $\Nv$. Although the trends appear to be linear, the investigation of further compounds would clarify the exact functional form of this dependence. Even though the ordering of the \CMG\ sample was not provided \cite{bel09}, it appears to fall on the B2 branch of the data.
(iii) The B2-ordered Co$_2$MnAl and CCFA have similar $\Nv$'s (7 and 6.95 valence electrons per atom) and very similar $D$ values.

The observations (i) -- (iii) suggest that $D$ is a function of both $\Nv$ and the crystallographic ordering. The latter observation is consistent with our previous study of a series of \CMS\ samples with a varying degree of L2$_1$ order \cite{gai08}. The determined $D$ values are shown in the inset of Fig.~\ref{f:D-Heuslers} and  linearly increase with increasing degree of L2$_1$ ordering. Extrapolating these data, it can be inferred that the $D$ value for a perfectly B2-ordered \CMS\  sample would be reduced to $\sim$56\% (\emph{i.e.}\ to $324$\,meV\,\AA$^2$) compared to an L2$_1$ ordered sample of the same composition (575\,meV\,\AA$^2$).  The extrapolated B2 value of $D$ for Co$_2$MnSi compares favorably with the experimental $D$ of B2-ordered \CFA\ (370\,meV\,\AA$^2$) (both \CMS\ and \CFA\ have $\Nv$ = 7.25\,\emph{e}$^-$/atom). It again suggests $D$ to be a function of $\Nv$ and the crystallographic ordering. 
Assuming, that $D_{B2}/D_{L2_1}\approx60\%$ for all Heusler compounds, the hypothetical $D$ values for L2$_1$ ordered CCFA and Co$_2$MnAl would rise to \emph{ca.} 330\,meV\,\AA$^2$ at RT. This value would however still be $\sim$70\,meV\,\AA$^2$ lower than the extrapolated behavior for the L2$_1$ branch %
\footnote{Another effect which may contribute to the small $D$ values stems from the fact that the measurement were done at RT. Assuming a $T$-dependent $D$ as in Ref.~\cite{dol98}, $D$ for CCFA and \CMA\ would be reduced by $\sim$6\% when measured at RT compared to their 0\,K values (their $T_C$'s are 750\,K, 630\,K, respectively \cite{blo03,web71}), yielding $D$ for 0\,K of $\sim$350\,meV\,\AA$^2$. Those corrected values are marked by empty squares in Fig.~\ref{f:D-Heuslers}(a). In the case of Co$_2$MnSi and Co$_2$FeSi (with $T_C$'s of 985\,K and 1100\,K, respectively \cite{web71,wur05}), the $D$ at RT would be reduced by only about 2\% with respect to their 0\,K value. Unfortunately, to the best of our knowledge, the $T$-dependence of $D$ has not been investigated so far for any Co$_2$-based Heusler compound. Finally, note that the $T_\mathrm{C}$'s of Co$_2$MnGe and Co$_2$FeAl are 905\,K \cite{web71} and $\sim$1000\,K \cite{fecher-private}.%
}.

One of the most attractive features of Heusler compounds is the possibility to tune their magnetic properties, such as $M_\mathrm{S}$ \cite{wur05,SP-Galanakis:PRB} and $T_\mathrm{C}$ \cite{kub07}, using the chemical handle provided by $\Nv$. As was discussed above, the exchange stiffness $D$ scales with $\Nv$, for a given atomic ordering. 
While a quantitative description of this trend is still elusive, here we point out a few features that help to understand the dependence of $D$ on $\Nv$.

\begin{figure}
\includegraphics[width=0.25\textwidth]{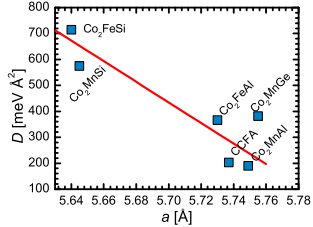}
\caption{%
\label{f:D_on_a}%
(color) Exchange stiffness $D$ as a function of the bulk lattice constant $a$. The solid line is a guide to the eye.
}
\end{figure}

In the following, we discuss three possible contributions which may give rise to the observed dependence between the composition (and $\Nv$) and the exchange integral $J$, which is related to the exchange stiffness $D$.  
(1) The increase of $\Nv$ adds electrons to the electronic structure. The additional electrons will be primarily found in the $t_{2g}$ orbitals of the transition metal \textsl{X} in the \CYZ\ Heusler compound \cite{kan07}. Thus, increasing $\Nv$ increases the electron density that will participate in exchange. Additionally, \textit{ab-initio} atom-resolved calculations of the density of states in \CYZ\ (\textsl{X}=Mn,Fe; \textsl{Z}=Al,Si) \cite{DOS-Fe,DOS-Mn,kub07} show that for a given non-magnetic element \textsl{Z}, substituting Mn by the more electronegative Fe results in a lowering of the energy of the \textsl{X}-based bands. This results in an improved alignment of the band energy between \textsl{X} and Co atoms, providing a stronger overlap of the electronic functions, and hence a stronger exchange interaction. 
(2) The difference in electronegativity between Co and Si ($-0.02$ on the Pauling electronegativity scale) is much less than between Co and Al ($0.27$). As such, Co-Si bonding is of more covalent character than Co-Al bonding \cite{kan07}. While this will impact the electronic properties, it is not clear how this affects the exchange interactions. However, this appears to be correlated to an enhanced exchange stiffness.
(3) Finally, increasing $\Nv$ is related with smaller atomic diameter, leading to 
a contraction of the unit cell due to the reduction in size of the constituent elements.  Hence, a stronger exchange interaction is expected due to the better overlap of orbitals, a result of the closer proximity between the magnetic elements. Indeed, larger exchange stiffnesses are generally observed for smaller (bulk) lattice constants~$a$, as is shown in Fig.~\ref{f:D_on_a}.

It is important to note that the trend between $D$ and $\Nv$ is not solely due to $\Nv$, as is the case for the magnetic moment determined by the Slater-Pauling rule \cite{SP-Galanakis:PRB}. In particular, the same exchange stiffness would not be expected \emph{a priori} for systems having the same $\Nv$, due to generally different electronic structures, and thus exchange integrals. However, we have observed the same exchange for \CMA\  and CCFA (having nearly the same $\Nv=7$ and 6.95\,\emph{e}$^-$/atom, respectively). Furthermore, the exchange value of \CMS\ extrapolated to the B2-ordered state is comparable to the $D$ values for \CFA\ and \CMG, all having $\Nv=7.25\,$\emph{e}$^-$/atom.

\begin{figure}[!h]
\includegraphics[width=0.37\textwidth]{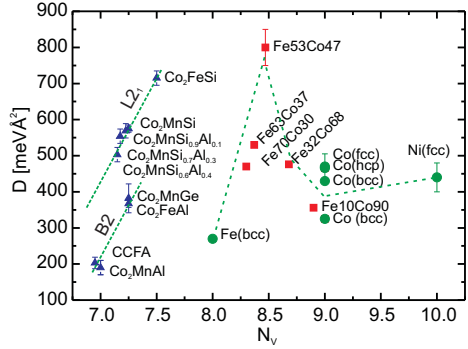}
\caption{%
\label{f:exch}%
(color) Exchange stiffness $D$ as a function of number of valence electrons $\Nv$  for \CYZ\ Heusler compounds, the ferromagnetic 3$d$-metals, and bcc-\FeCo\ compounds. Lines are guides to the eye.
}
\end{figure}

Figure~\ref{f:exch} compares the exchange stiffness $D$ associated to a variety of Co$_2$-based Heusler compounds, the ferromagnetic 3$d$-metals, and Fe-rich bcc-\FeCo\ intermetallic compounds, as a function of $\Nv$. Note that these are experimental  data collected from our own work on Co$_2$-based Heuslers \cite{ham09cms,gai09ccfa,gai08,kub09} and various third-party  publications \cite{liu94, liu96, web71,fuk06, rez03, shi68, bel09,sha05,blo03,wur05}. 

As discussed above, Co$_2$-based Heusler compounds appear to follow a roughly linear dependence of $D$ on $\Nv$, reaching a maximum value with \CFS\ ($D=715\pm20$\,meV\,\AA$^2$).  
The \FeCo\ compounds also provide a roughly linear dependence between $D$ and $\Nv$ in the $x$ range of 50--100, reaching a maximum value of  $D=800\pm50$\,meV\,\AA$^2$ with Fe$_{53}$Co$_{47}$ \cite{liu94} which is (to the best of our knowledge), the largest exchange stiffness reported. This shows that the value we find for \CFS\ is extraordinarily large. It is larger than $D$ of the pure ferromagnetic 3$d$ metals, and nearly as large as the maximum value of $D$ obtained for the \FeCo\ series. Such an extraordinary value of \CFS\ is surprising, particularly when taking into account that one quarter of the constituent atoms is a~non-magnetic element (Si). It is also worth noting that the exchange constant $A$ of \CFS\ (31.5\,pJ/m) is also only surpassed by Fe$_{53}$Co$_{47}$ (38.4\,pJ/m), and is on par (within experimental error) with Fe$_{63}$Co$_{37}$ (32.1\,pJ/m) \cite{liu94}.

In conclusion, we have found an extraordinary large exchange stiffness $D$ in \CFS, which is nearly as high as the record exchange of Fe$_{53}$Co$_{47}$. 
The careful comparison between exchange stiffnesses in various Co$_2$-based Heusler compounds shows that the exchange stiffness $D$ is a function of the number of valence electrons $\Nv$ and the crystallographic order. Hence, the dependence of $D$ on $\Nv$ is provided by two different branches for L2$_1$ and B2 order,
the latter one being about 60\% of the former. 
In both branches, $D$  increases with increasing  $\Nv$, which corresponds to the larger overlap of the wave functions due to (i) an increase of the electron density, (ii) better match of the band energies and (iii) shrinking of the lattice constant.
The establishment of such trends is crucial towards devising new materials, as well as providing guidelines towards a better first-principle understanding of the underlying electronic structures of cobalt-Heusler compounds.

This project was financially supported by the DFG Research Unit 559. S.T. gratefully acknowledges the Alexander von Humboldt foundation for a PDF.
We thank J. K\"ubler, H.J. Elmers, J. Kudrnovsk\'y, and G. Fecher for stimulating discussions.
We thank T.~Kubota for providing data on Co$_2$MnAl$_x$Si$_{1-x}$ prior to publication.


\begin{thebibliography}{10}

\bibitem{sak06}
Y. Sakuraba, M. Hattori, M. Oogane, Y. Ando, H. Kato, A. Sakuma, T. Miyazaki,
  and H. Kubota, Appl. Phys. Lett. {\bf 88},  192508  (2006).

\bibitem{review:Inomata}
K. Inomata, N. Ikeda, N. Tezuka, R. Goto, S. Sugimoto, M. Wojcik, and E.
  Jedryka, Sci. Technol. Adv. Mater. {\bf 9},  014101  (2008), and references
  therein.

\bibitem{Skomski}
P.~A. Dowben and R. Skomski, J. Appl. Phys. {\bf 95},  7453  (2004).

\bibitem{rajanikanth:063916}
A. Rajanikanth, Y.~K. Takahashi, and K. Hono, J. Appl. Phys. {\bf 105},  063916
   (2009), and references therein.

\bibitem{Thoene}
J. Thoene, S. Chadov, G. Fecher, C. Felser, and J. K\"{u}bler, J. Phys. D:
  Appl. Phys. {\bf 42},  084013  (2009), and references therein.

\bibitem{gai08}
O. Gaier, J. Hamrle, S.~J. Hermsdoerfer, H. Schulthei\ss\, B. Hillebrands, Y.
  Sakuraba, M. Oogane, and Y. Ando, J. Appl. Phys. {\bf 103},  103910  (2008).

\bibitem{ham09cms}
J. Hamrle, O. Gaier, S.-G. Min, B. Hillebrands, Y. Sakuraba, and Y. Ando, J.
  Phys. D: Appl. Phys. {\bf 42},  084005  (2009).

\bibitem{gai09ccfa}
O. Gaier, J. Hamrle, S. Trudel, A.~C. Parra, B. Hillebrands, C.~H. E.~Arbelo,
  and M. Jourdan, J. Phys. D: Appl. Phys. {\bf 42},  084004  (2009).

\bibitem{liu94}
X. Liu, R. Sooryakumar, C.~J. Gutierrez, and G.~A. Prinz, J. Appl. Phys. {\bf
  75},  7021  (1994).

\bibitem{schneider2009}
H. Schneider, E. Vilanova, B. Balke, C. Felser, and G. Jakob, J. Phys. D: Appl.
  Phys. {\bf 42},  084012  (2009).

\bibitem{hil99}
B. Hillebrands, Rev. Scien. Instr. {\bf 70},  1589  (1999).

\bibitem{Hilleb-LSiS}
B. Hillebrands,  in {\em Light Scattering in Solids VII}, edited by M. Cardona
  and G. G\"{u}ntherodt (Springer-Verlag, Heidelberg, 2000).

\bibitem{hil90}
B. Hillebrands, Phys. Rev. B {\bf 41},  530  (1990).

\bibitem{SP-Galanakis}
I. Galanakis, P. Mavropoulos, and P.~H. Dederichs, J. Phys. D: Appl. Phys. {\bf
  39},  765  (2006).

\bibitem{fech06}
G.~H. Fecher, H.~C. Kandpal, S. Wurmehl, C. Felser, and G. Sch\"{o}nhense, J.
  Appl. Phys. {\bf 99},  08J106  (2006).

\bibitem{kueb84}
J. K\"ubler, Physica B and C {\bf 127},  257  (1984).

\bibitem{kub07}
J. K\"{u}bler, G.~H. Fecher, and C. Felser, Phys. Rev. B {\bf 76},  024414
  (2007).

\bibitem{kub09}
T. Kubota and \textit{et al} (unpublished).

\bibitem{bel09}
M. Belmeguenai, F. Zighem, Y. Roussign\'{e}, S.-M. Ch\'{e}rif, P. Moch, K.
  Westerholt, G. Woltersdorf, and G. Bayreuther, Phys. Rev. B {\bf 79},  024419
   (2009).

\bibitem{wur05}
S. Wurmehl, G.~H. Fecher, H.~C. Kandpal, V. Ksenofontov, C. Felser, H.-J. Lin,
  and J. Morais, Phys. Rev. B {\bf 72},  184434  (2005).

\bibitem{SP-Galanakis:PRB}
I. Galanakis, P.~H. Dederichs, and N. Papanikolaou, Phys. Rev. B {\bf 66},
  174429  (2002).

\bibitem{kan07}
H.~C. Kandpal, G.~H. Fecher, and C. Felser, J. Phys. D: Appl. Phys. {\bf 40},
  1507  (2007).

\bibitem{DOS-Fe}
K. \"{O}zdo\u{g}an, B. Akta\c{s}, I. Galanakis, and E. \c{S}a\c{s}io\u{g}lu, J.
  Appl. Phys. {\bf 101},  073910  (2007).

\bibitem{DOS-Mn}
K. \"{O}zdo\u{g}an, E. \c{S}a\c{s}io\u{g}lu, B. Akta\c{s}, and I. Galanakis,
  Phys. Rev. B {\bf 74},  172412  (2006).

\bibitem{liu96}
X. Liu, M.~M. Steiner, R. Sooryakumar, G.~A. Prinz, R.~F.~C. Farrow, and G.
  Harp, Phys. Rev. B {\bf 53},  12166  (1996).

\bibitem{web71}
P.~J. Webster, J. Phys. Chem. Solids {\bf 32},  1221  (1971).

\bibitem{fuk06}
T. Fukuda, M. Yuge, T. Terai, and T. Kakeshita, Journal of Physics: Conference
  Series {\bf 51},  307  (2006).

\bibitem{rez03}
S.~M. Rezende, M.~A. Lucena, A. Azevedo, F.~M. de~Aguiar, J.~R. Fermin, and
  S.~S.~P. Parkin, J. Appl. Phys. {\bf 93},  7717  (2003).

\bibitem{shi68}
G. Shirane, V.~J. Minkiewicz, and R. Nathans, J. Appl. Phys. {\bf 39},  383
  (1968).

\bibitem{sha05}
S. Shallcross, A.~E. Kissavos, V. Meded, and A.~V. Ruban, Phys. Rev. B {\bf
  72},  104437  (2005).

\bibitem{blo03}
T. Block, C. Felser, G. Jakob, J. Ensling, B. M\"uhling, P. G\"utlich, and R.
  Cava, J. Sol. Stat. Chem. {\bf 176},  646  (2003).

\bibitem{dol98}
L. Vasiliu-Doloc, J.~W. Lynn, A.~H. Moudden, A.~M. de~Leon-Guevara, and A.
  Revcolevschi, Phys. Rev. B {\bf 58},  14913  (1998).

\bibitem{fecher-private}
G. Fecher, private communication.

\end{thebibliography}

\end{document}